\documentclass[submission,copyright,creativecommons]{eptcs}
\providecommand{\event}{ACT 2019} 

\usepackage{graphicx}
\usepackage{amsmath}
\usepackage{amssymb}
\usepackage{tikz-cd}
\usetikzlibrary{cd}
\usepackage{jlcode}
\usepackage{hyperref}
\usepackage{subcaption}
\usepackage[section]{placeins}
\usepackage{hyphenat}

\newcommand{\semanticModels}{\nohyphens{\texttt{SemanticModels.jl}}}
\newcommand{\julia}{\path{julia}}

\title{A Compositional Framework for Scientific Model Augmentation}
\author{Micah Halter \qquad\qquad\qquad Christine Herlihy \qquad\qquad James Fairbanks
\institute{Georgia Tech Research Institute\\ Atlanta, USA}
\email{Micah.Halter@gtri.gatech.edu \qquad\qquad\qquad\qquad\qquad\qquad\qquad James.Fairbanks@gtri.gatech.edu}
}
\def\titlerunning{A Compositional Framework for Scientific Model Augmentation}
\def\authorrunning{M. Halter, C. Herlihy \& J. Fairbanks}
\begin{document}
\maketitle

\begin{abstract}
Scientists construct and analyze computational models to understand the world.
That understanding comes from efforts to augment, combine, and compare models of related phenomena.
We propose \semanticModels{}, a system that leverages techniques from static and dynamic program analysis to process executable versions of scientific models to perform such metamodeling tasks.
By framing these metamodeling tasks as metaprogramming problems, \semanticModels{} enables writing programs that generate and expand models.
To this end, we present a category theory-based framework for defining metamodeling tasks, and extracting semantic information from model implementations, and show how this framework can be used to enhance scientific workflows in a working case study.
\end{abstract}


\semanticModels{} facilitates several metamodeling tasks by detecting and exploiting the implicit relationships between the semantically rich, natural language-based representations of scientific knowledge found in academic papers, and the relatively semantically sparse, but modular, precise representations found in code.
We represent this knowledge with categories and support metamodeling tasks that may be exploratory, iterative, and/or inter-disciplinary in nature.
Our theoretical contributions form the basis for analyzing the metamodeling tasks scientists perform, and our software can augment scientific workflows to assist scientists in their day to day work.

\paragraph{Motivation}
Progress in science often comes from adapting and extending models from prior work(s) to address new problems, but current scientific research workflows make leveraging components from existing workflows difficult.
This is due in part to the fact that modern scientific inquiry often lends itself to highly tailored, procedural scripts that are primarily intended to produce and record results, with less attention on software engineering best practices~\cite{deelman2017, wilson2014}.

Scientific code contains a large amount of sophisticated domain knowledge that is known to the author(s), but is not explicitly represented in the code, and is therefore not always clear to a reader or user.
Such semantic modeling information includes principles, rules, and constraints imposed by the physical phenomena being modeled.
For example: (1) stochastic systems are modeled with probability values, which are constrained to be between 0 and 1; (2) physical measurements have units and must obey the laws of dimensional analysis, which prohibits computation such as $3m + 4m/s$; and (3) signal processing algorithms must treat time domain and frequency domain signals differently, even though they are both represented by arrays of floating point numbers. This work is grounded in the belief that a framework that augments scientific workflows without requiring a wholesale reimplementation in a domain specific language is most effective for real world application by scientists.

\paragraph{Significance}
As computational models of physical, biological, and engineered systems grow increasingly sophisticated, program analysis tools must be able to understand and manipulate these models.
We introduce a formalism to study the augmentation of scientific modeling code.
These ideas are implemented in a software package for analyzing and manipulating models written in the Julia programming language~\cite{doi:10.1137/141000671}.
The Julia language is ideal for this problem because it is widely used in scientific computing and includes a capable type system  with multiple dispatch.
The Julia type system can encode information about model semantics so that the compiler can understand, enforce, and manipulate these semantics.
These manipulations are studied in the context of epidemiological models, but are broadly applicable to both agent-based and differential equation-driven simulations, as well as statistical and machine learning models.

\hypertarget{{related_works}}{%
\paragraph{Related Work}\label{{related_works}}}

The semantic metamodeling system proposed here is informed by foundational concepts from several disciplines, including software engineering, programming language theory, natural language processing, and statistical meta-analysis. Software engineering emphasizes modular design, automation of repeated tasks, and incremental software modifications~\cite{BOEHM19833}.
Within this context, refactoring code refers to a process in which developers modify portions of existing code to preserve or improve correctness while increasing maintainability.
The correctness of a piece of software can be formally defined, and it is possible to design an automatic verification system; this is a well-established field of study within theoretical computer science~\cite{Hoare:1969:ABC:363235.363259}.

Our system extends this notion of automatically verification of program correctness to the semantic level by identifying, connecting, and verifying the unwritten invariants of scientific modeling code.
In contrast to the explicit type and syntax rules employed by traditional verification approaches, the rules we seek to identify, extract, and use are often informally specified or encoded in non-operational code and text, such as documentation, comments, and/or variable naming conventions.
An analogous framework based on using categories to represent semantic information from data science workflows is based on ontologies and dataflow graphs~\cite{DBLP:journals/corr/abs-1807-05691}.
Categorification is used to provide unity and depth of understanding to existing scientific modeling frameworks~\cite{baez2017}. The categorification of individual modeling frameworks exposes the structure of the scientific models in a domain and enables compositionality of sciences.
Our software implementation is intended to serve as a generic framework for implementing categorified modeling frameworks by \emph{post hoc} grafting a more elegant framework onto code implemented with pre-categorification libraries.

Disciplines of science and applied mathematics are amenable to formal systems of semantic verification; for example, unitful computation can be verified via dimensional analysis~\cite{drobot1953}, while linear algebra can be verified by matrix dimension checks~\cite{macedo2013}.
Both are understandable through static program analysis and categorification.
We believe that categorification enables the formalization of many informal systems of scientific modeling with type systems, providing the computational infrastructure for efficient and elegant implementations.


\paragraph{Theoretical Foundations}
We begin by presenting a formal framework that can be used to represent and reason about these different metamodeling use cases.
We then provide examples from epidemiology to illustrate how this framework, and the associated semantic knowledge graph construction process, can be applied to augment real-world scientific workflows~\cite{epirecipes}.
Our motivation for focusing on epidemiology is twofold: the associated literature demonstrates the use of a shared model structure with many variations.
Furthermore, the math represented therein spans both discrete and continuous systems of equations, and is solved by a diverse set of algorithms.

Scientific programmers represent models at three levels: (1) as a set of domain concepts understood by the developer, but not explicitly stated or encoded; (2) as code implementations in a high-level language; and (3) as an executable program compiled or interpreted on a specific computer system.
We study representations of programs that span level 1 and 2 as categories that represent the knowledge of scientists as expressed in code.

\begin{figure}[hbtp]
\centering
\includegraphics[width=0.9\textwidth]{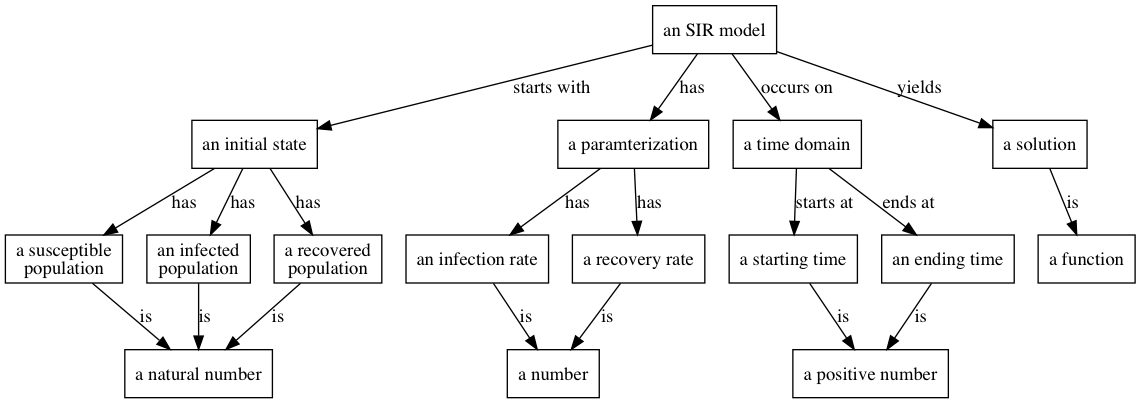}
\caption{Ologs can be used to represent the structure of scientific models without the mathematics. Objects are represented by nouns and relationships between objects are represented with verbs.}\label{fig:olog}
\end{figure}

\begin{figure}[tbh!]
\begin{center}
\begin{subfigure}[tbh!]{0.49\textwidth}
\includegraphics[width=\textwidth]{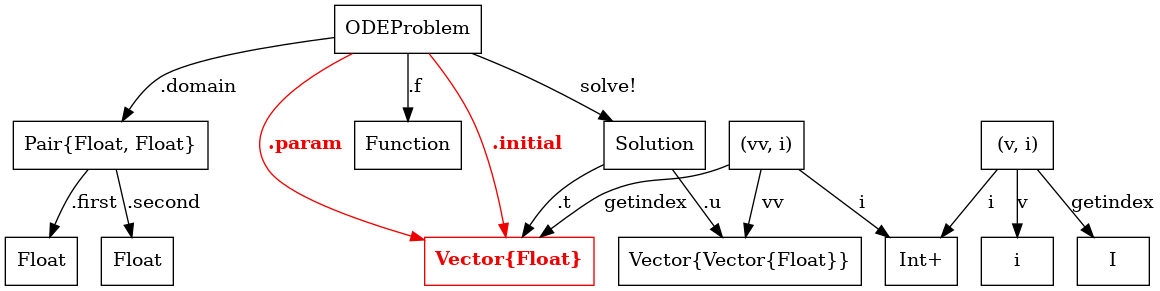}
\caption{}
\label{fig:type_ambig}
\end{subfigure}
\begin{subfigure}[hbt]{0.49\textwidth}
\includegraphics[width=\textwidth]{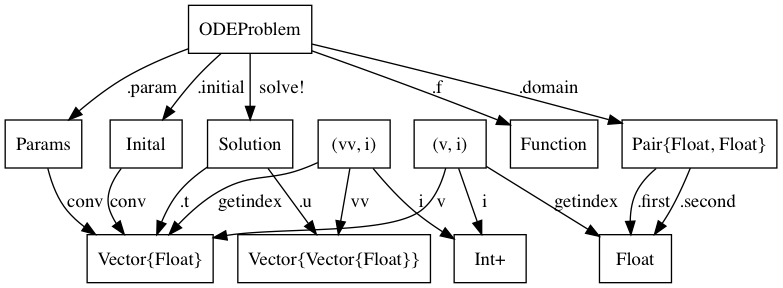}
\caption{}
\label{fig:type_unambig}
\end{subfigure}
\caption{\ref{fig:type_ambig} Type diagram of the SIR model. The red arrows and boxes show that \texttt{Vector\{Float\}} is reused for two semantically different types. \ref{fig:type_unambig} Type diagram of the SIR model with ambiguity resolved by introducing the \texttt{Params} and \texttt{Initial} types. Some type names are abbreviated for clarity i) Int, v) Vector\{Int\}, and vv) Vector\{Vector\{Int\}\}. Functions names starting with a \texttt{"."} are struct field lookups in Julia.}
\label{fig:modeltypes}
\end{center}
\end{figure}

A model \(M=(D,R,f)\) is a tuple containing a set \(D\), the domain, and a set \(R\), the co-domain, with a function \(f:D\mapsto R\).
If \(D\) is the cross product of sets \(D_1 \times D_2 \cdots D_k\), then \(f = f(x_1\dots x_k)\), where \(x\) are the independent variables of \(M\).
If \(R=R_1 \times R_2 \cdots R_d\), then \(R_i\) are the dependent variables of \(M\).
What is intuitively \emph{the same model} can be represented in several different categories.
Each representation considers different aspects of the model's structure.
For example, Figure~\ref{fig:olog} shows the SIR model in the category of ontology logs (ologs), and Figure~\ref{fig:type_ambig} shows the same model in the category of programs.

The parsimonious formalization of what constitutes a valid transformation, or set of rules for modifying or combining models, requires us to assess not only mathematical and programmatic behavior of the system, but also the extent to which the resulting set of models are internally consistent and reflective of domain-specific scientific facts.
We address this with multiple representations of program semantics that capture different types of model structure.

A Category $C$ is a set of objects and morphisms, which are structure-preserving functions between the objects.
Common examples of categories include the category of all groups, the set of all finite graphs, and the set of all finite preorders~\cite{Spivak:2014:CTS:2628001}.
Ologs are a diagrammatic approach to formalizing scientific knowledge used to precisely specify a conceptual model of a phenomenon or experiment~\cite{10.1371/journal.pone.0024274}.
An olog is composed of types (boxes) and aspects (edges).
Figure~\ref{fig:olog} represents the susceptible-infected-recovered (SIR) model as an olog.

All programs in a strongly typed language have a set of types and functions that map values between those types.
For example, the Julia program: \texttt{a = 5.0; b = 1; c = 2*a; d = b + c;} has the types \texttt{\{Int, Float\}} and functions \texttt{\{*, +\}}, which are both binary functions.
These programs can be represented as a category, where the objects are the types and the morphisms are the functions.
We refer to the input type of a function as the domain and the output type as the codomain of the function.
Multi-argument functions are represented with tuple types representing their arguments.
For example\footnote{The \texttt{a::A} operator in Julia asserts that the value of \texttt{a} is an instance of type \texttt{A}} \texttt{{+}(a::Int,b::Int)::Int} is a function ${+}: Int\times Int \to Int$.
These type categories are well studied in the field of functional programming.
We apply these categories to the study of mathematical models.
There is a spectrum between conceptual knowledge and compiler knowledge, with ologs lying toward the conceptual end and compiler dataflow graphs at the programmatic end.

Functional programming and category theory are intertwined and base the analysis of programs on the types and functions used in the program~\cite{Wadler:1992:EFP:143165.143169}.
\semanticModels{} implements a dynamic analysis tool to extract the runtime type information for every function.
That is, to build a graph where the nodes are types and the edges are functions, where a function $f$ connects types $T,U$ if $T$ is the type of $f$'s arguments and $U$ is the type of $f$'s output values as expressed in \julia{} syntax, \texttt{f(x::T)::U}.
This theoretic approach enables reasoning over the semantics of programs.

The most salient consequence of programming language theory is that the more information a programmer can encode in the type system, the more helpful the programming language can be for improving performance, quality, and correctness.
Haskell programmers often use the type system to encode program semantics to improve software quality~\cite{manzino2014}.
\semanticModels{} uses the type system to encode model semantics to improve understanding, adaptability, and extensibility of the modeling code.
Category theory provides a natural vehicle for expressing this information.

\paragraph{Semantic Integrity of Modeling Programs}
Model developers use conventions to encode semantic constraints into their code -- for example, prefacing all variables that refer to time with a \texttt{t\_}, such as \texttt{t\_start, t\_end}.
This semantic constraint that all variables named \texttt{t\_} are temporal variables is not encoded in the type system.
Behavioral subtypes are one way of encoding such information, but they are not widely used in scientific computing.
Another example is that vectors of different lengths are incompatible in the context of arithmetic operations.
In a compartmental model, the number of initial conditions must match the number of compartments.
For example in an SIR model, there are 3 initial conditions, $[S,I,R]$, and there are 2 parameters $[\beta, \gamma]$.
Computational systems employed by scientists will use a runtime check on dimensions to detect malformed linear algebra~\footnote{Julia, Scientific Python, and Matlab use run time checks, the C++ library \href{https://eigen.tuxfamily.org/dox/group__TutorialMatrixArithmetic.html}{Eigen} supports both static and dynamic dimension verification}.
Scientists rely on this limited form of semantic integrity checking provided by the language.
In contrast, \semanticModels{} is intended to rigorously apply such integrity checking across the modeling ecosystem.

Our goal is to encode the maximum amount of information from scientific codes into the type system, where algorithms can analyze the integrity of programs in the language of categories.
For example, if there are types $S,T$ and functions $f,g: S\rightarrow T$ such that $Codom(f) = Codom(g)$ but $Range(f) \cap Range(g) = \emptyset$, then we say that the types are \emph{ambiguous}.
In order to more fully encode program semantics into the type system, the programmer (or an automated system) should introduce new types into the program to represent these disjoint subsets.
Category theory shows both why this is a problem for program analysis\footnote{If model transformations are represented as functors in this category, this form of ambiguity prevents the type system from enforcing semantic correctness of model transformations} and how to solve it with \emph{union types}.

Returning to the SIR model example, Figure~\ref{fig:modeltypes} shows how the \texttt{.param} and \texttt{.initial} functions both map \texttt{Problem} to \texttt{Vector\{Float\}} but with disjoint ranges.
The mathematics of the model dictate that parameters and initial conditions have different dimensions and are thus incompatible vectors.
Any program analysis of the model will be hampered by the ambiguity introduced by using the same type to represent two different concepts.
The functions \texttt{.first} and \texttt{.second}, which provide the beginning and end of the time domain of the system, have overlapping ranges and are comparable as times.
This is an example of how programming language ideas can improve the analysis of computational models.

\paragraph{Program Analysis of Models} Static program analysis provides direct access to the function call graph; however, inferred types and runtime values require dynamic analysis.
To do this, we inject metadata collection statements into each program's AST, so that when the AST is evaluated, we are able to dynamically capture variable assignments and function calls.
In practice, scientists prefer dynamic languages that facilitate faster development, but they are challenging for static analysis techniques.
Julia provides a hybrid of static compilation and dynamic execution that is amenable to rapid development, prototyping, and program analysis.

\begin{figure}[bth]
\centering
\begin{subfigure}{0.4\textwidth}
\centering
$ f(a,b) = 2\cdot(a * b) $
\end{subfigure}~
\begin{subfigure}{0.6\textwidth}\centering
$ g(a,b) = 2\cdot(a/b) $
\end{subfigure}
\begin{subfigure}{0.4\textwidth}\centering
\begin{tikzcd}
  Z\times Z \arrow[d, "*", olive] \arrow[d, bend right, "\pi_1"] \arrow[d, bend left, "\pi_2"]
  \\ Z \arrow[loop right, "2\times", blue]
 \end{tikzcd}
\end{subfigure}~
\begin{subfigure}{0.59\textwidth}\centering
\begin{tikzcd}
  Z\times Z \arrow[r, "/", olive] \arrow[d, bend right, "\pi_1"] \arrow[d, bend left, "\pi_2"]
    & R\oplus \texttt{Error} \arrow[loop right, "2\times", blue] \\
  D
\end{tikzcd}
\end{subfigure}
\caption{Programs with the same structure admit a fully faithful functor from one to the other. In this case the functor is shown with color identifying which functions are mapped to each other. The objects are types and the morphisms are functions (subroutines). The morphisms $\pi_i$ represent the projection functions that select the $i$th element from a tuple.}\label{fig:arithcat}
\end{figure}

Model augmentation refers to the set of metamodeling programs where a scientist takes a model, \(M\), and a transformation, \(T\), and uses the transformation to construct a new model, \(T(M)\).
In complex, high performance modeling and simulation software, these changes can be very labor intensive.
To facilitate the identification of program components that are good candidates for modification, \semanticModels{} provides a bundle of tools that rely on Julia metaprogramming (e.g., expression manipulation via Lisp-style macros) to modify programs for the purpose of dynamic information extraction.
Model augmentation is implemented via program transformations. Figure~\ref{fig:arithcat} illustrates how programs can be represented as categories and transformations between those programs are functors. Many families of mathematical and scientific models have a separation between structure and values.
For example, in dynamical systems and reaction networks, there is the structure of the equations and then the specific rate parameters.
Representing programs as cateogries and transformations can exploit this notion of ``same structure, but different values'' across a wide class of models.
The simple example in Figure~\ref{fig:arithcat} shows how properties of the functor between two models tell use about the relationship between those two models.

We can illustrate another example of this in a simple yet realistic agent-based model (ABM).
For example, an SIR model can be implemented as an ABM, as depicted in Figure~\ref{src:abm}.
This script defines a basic agent-based model of disease spread called SIRS.
Each agent is in one of 3 states: $S$ Susceptible, $I$ Infected, $R$ Recovered, and the agents transition between states.

By viewing model transformations as functors between model categories, we can implement model transformations that preserve structure while changing the behavior of the model (e.g. by adding or removing capabilities).
Our system includes the capability to add and remove states and change behaviors of the model, while capturing the nature of this change in a data structure, allowing users to probe the relationship between a set of models.
For example, the SIRS model does not have the structure necessary to model a fatal disease as there is no $D$ component, so a scientist would have to change the code.
By representing these changes as model transformations, we are able to add a new state, $D$ Dead, and the necessary transitions to enable the modeling of fatal diseases.

\begin{figure}[htb!]
\begin{lstlisting}
import Base: count
abstract type AgentModel end

# the main data structure
mutable struct StateModel <: AgentModel
    states
    agents
    transitions
    loads
end

function main(nsteps)
    # initialize the model
    n = 20
    a = fill(:S, n)
    ρ = min(1, max(0, 0.5 + randn(Float64)/4)) # chance of recovery
    μ = 0.5 # chance of immunity
    T = Dict(
        :S=>(x...)->rand(Float64) < stateload(x[1], :I) ? :I : :S,
        :I=>(x...)->rand(Float64) < ρ ? :I : :R,
        :R=>(x...)->rand(Float64) < μ ? :R : :S,
    )
    sam = StateModel([:S, :I, :R], a, T, zeros(Float64,3))
    # advance the simulation by nsteps steps
    newsam = step!(deepcopy(sam), nsteps)
    # compute summary statistics from the simulation
    counts = describe(newsam)
    return newsam, counts
end

end
\end{lstlisting}
\caption{A Simple ABM for a SIR modeling. The agents go from $S\mapsto I\mapsto R\mapsto S$ based on random numbers.
The probability of $S\mapsto I$ is dependent on the fraction of agents in state $I$.
The probability of recovering is a constant $\rho$, and the disease confers some temporary immunity with probability $\mu$.
The function definitions for \texttt{step!}, which advances the timesteps of the ABM, and \texttt{describe}, which computes summary statistics for downstream analysis, are omitted for space.}\label{src:abm}

\end{figure}

\begin{figure}[htbp]
\begin{subfigure}{.65\textwidth}
\includegraphics[width=0.98\textwidth]{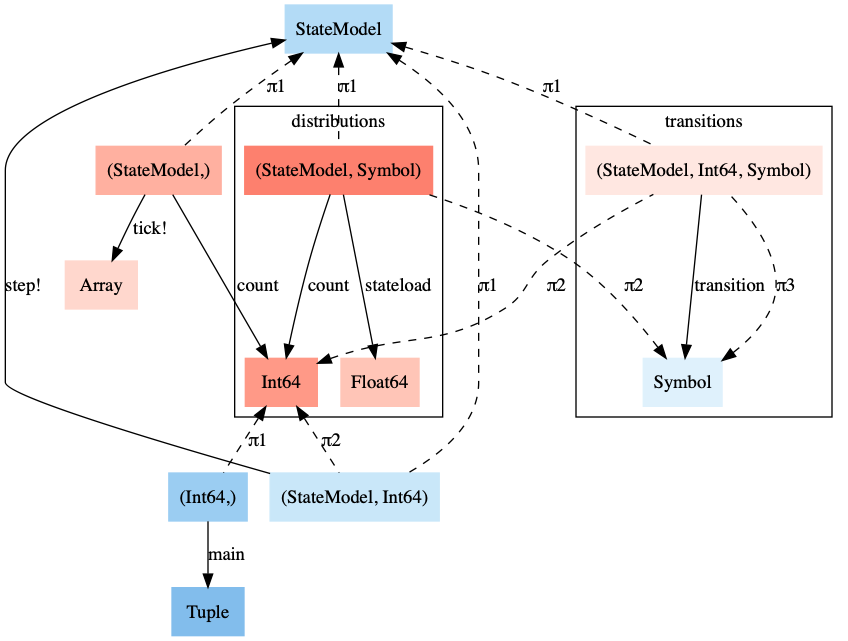}
\caption{}\label{fig:typegrapha}
\end{subfigure}~
\begin{subfigure}{.32\textwidth}
\includegraphics[width=0.98\textwidth]{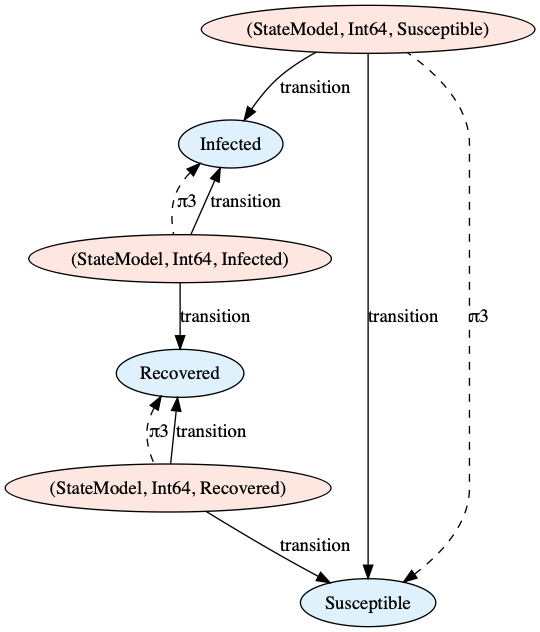}
\caption{}\label{fig:typegraphb}
\end{subfigure}

\begin{subfigure}{.85\textwidth}
\includegraphics[width=0.98\textwidth]{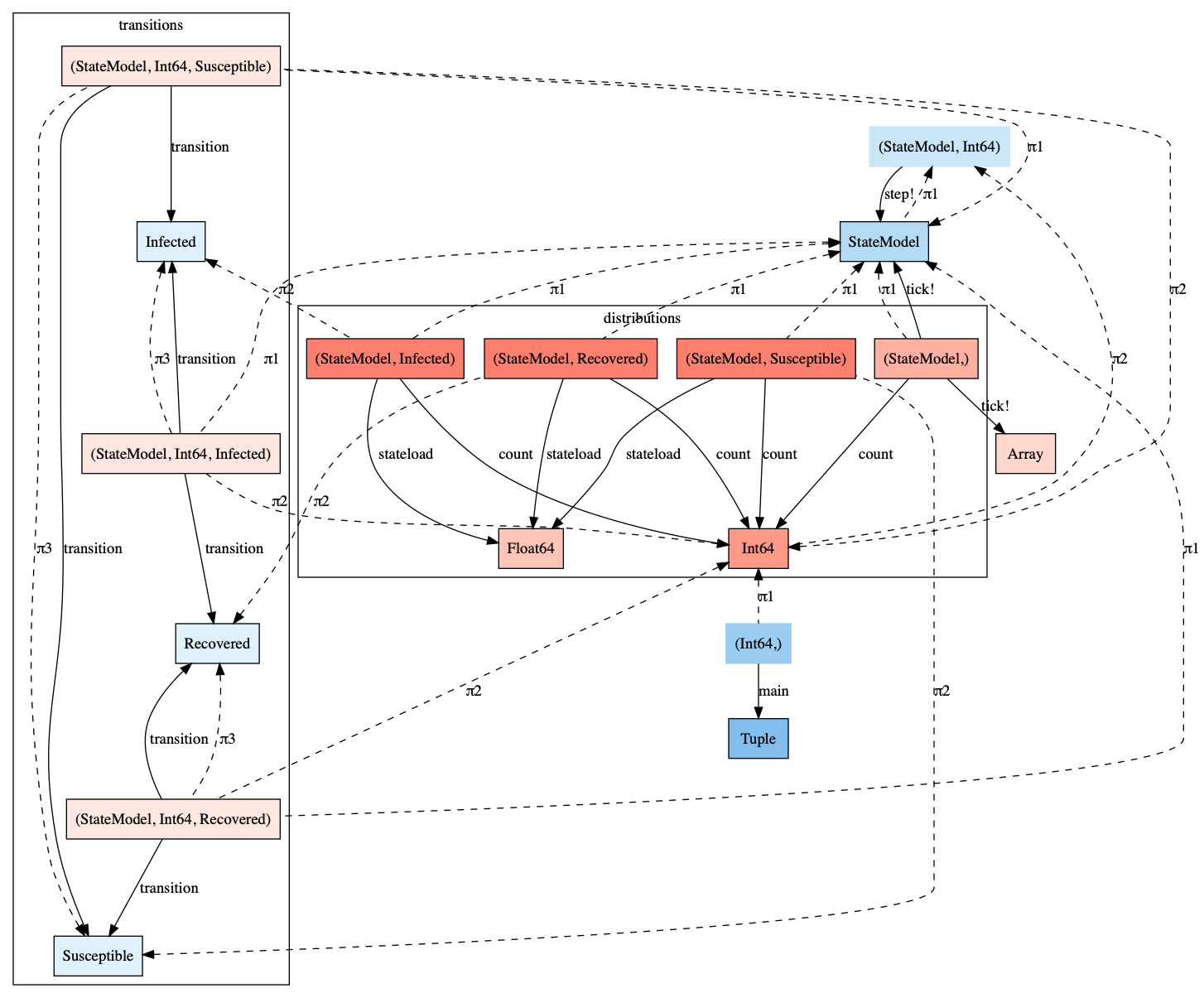}
\caption{}\label{fig:typegraphc}
\end{subfigure}~
\caption{\ref{fig:typegrapha} The input typegraph of the code corresponding to Figure~\ref{src:abm}.
\ref{fig:typegraphb} The states of the agents as expressed in a typegraph.
\ref{fig:typegraphc} The typegraph of the code refactored to encode model semantics into the type system. Note  $S,I,R$ are represented by the singleton types $Susceptible, Infected, Recovered$ respectively, and the colors show a graph homomorphism between the type graphs. The edges showing projection functions are dotted for visual clarity in the diagram, and the boxes labeled transition and distributions are identified as meaningful subgraphs of the program representing the agent-state transitions, and the calculation of population distribution respectively. }\label{fig:typegraph}
\end{figure}

Another type of model transformation includes refactoring models to introduce more structure into the program that can be exploited by program analysis techniques in the compiler.
One can transform the model shown in Figure~\ref{src:abm} by re-factoring the Symbol representation of the states into singleton types, and naming the anonymous functions as \texttt{transition}.
Figure~\ref{fig:typegraph} illustrates how a program transformation aimed at encoding model semantics into the type system can change the typegraph of the program.
The colors in Figure~\ref{fig:typegraph} show a graph homomorphism between the type graphs.
This homomorphism maps concepts from the new model to the old model.
This example illustrates another point: while we naturally think of \emph{model augmentation} as turning a simple model into a more complex model; when applying category reasoning, it is more convenient to think of a transformation $m' = t(m)$ as inducing a functor $\phi:m' \mapsto m$.
This functor, $\phi:m' \mapsto m$, takes a complex model and simplifies it, which provides an interpretation of $m'$ in terms of $m$. 

Subsequently, we can also see how models can be represented in different categories, and the notions of category theory in those different contexts allow for different kinds of analysis of the models.
Specifically, Figure~\ref{fig:typegraph} shows a pair of models represented as graphs and a functor between them (graph homomorphism) that relate the types in one model to the types in the other model.
The nature of the functor is determined by the category used and the models' structure.

The program transformation shown in Figure~\ref{fig:typegraph} is best viewed as a refactoring, where the maintainability and robustness of a model was improved without changing its behavior as a mathematical function.
A typical model augmentation does change the behavior of the model as a mathematical function in order to add capabilities or adapt the model to a new physical phenomena.
Given an algorithmic mechanism for changing scientific models, the first question of any scientist will be, ``Which model should I use?'' This question is answered by model selection.

\FloatBarrier
\paragraph{Algebraic Model Selection} When model families are parameterizable by $\mathbb{R}^n$, statistical theory can show how to choose the best model using a regularization process.
Take, for example, polynomial regression: where data is represented by random variable(s).
Define $X\in\mathbb{R}^{n\times d}$ as the independent variables and $y\in \mathbb{R}^n$ as the dependent (target) variable. Polynomial least squares regression solves the optimization problem $\min_p \|y - p(X) \|_2$,  where $p(X) = \sum_i \beta_i x^i$ and $\|\cdot\|_2$ is the two norm.

For model selection in polynomial regression, one must choose a polynomial degree and set of non-zero coefficients to define the model. Statisticians use the LASSO to select the best polynomial. LASSO is defined as $min_\beta \sum_j (\sum_i \beta_i x^i_j - y_j)^2 + \sum_i |\beta|$, where $j$ ranges over samples and $i$ ranges over polynomial terms.
Regularization generally works when the space of models can be parameterized by a continuous parameter, and the loss function can be modified to support the regularization penalty.

Meanwhile, for complex models, the space of possible models cannot be parameterized continuously.
Additionally, the inclusion of the regularization penalty often increases the complexity of the solver, because the difficulty of the optimization problem increases (eg, addition of LASSO regularization spoils the quadratic properties of the least squares problem).
Sparse polynomial regression (polynomials where the set of non-zero coefficients is small and known \emph{a priori}) is not continuously parameterizable and we need an algebraic perspective.

\begin{figure}[hbtp]
\centering
\footnotesize
\begin{subfigure}[b]{0.45\textwidth}
\begin{tikzcd}
  & & \emptyset \ar[dl, "T_x"]\ar[dr, "T_1"]& \\
  & T_x \ar[dl, "T_x"]\ar[d, "T_1"] & & T_1\ar[d, "T_x"]\ar[loop right, "T_1"] \\
   T_xT_x \ar[d, "T_1"]  & T_xT_1 \ar[d, "T_x"]\ar[loop right, "T_1"]&& T_1T_x \ar[dl, "T_x"]\ar[d, "T_1"] \\
   T_xT_xT_1 & T_xT_1T_x & T_1T_xT_x & T_1T_xT_1
\end{tikzcd}
\caption{}\label{fig:monoid_ab}
\end{subfigure}~
\begin{subfigure}[b]{0.45\textwidth}
\begin{tikzcd}
  && & 1\ar[d, "T_x"]\ar[loop right, "T_1"]& \\
  && & x\ar[dl, "T_x"]\ar[d, "T_1"]& \\
  && x^2 \ar[dl, "T_x"]\ar[d, "T_1"] & x+1\ar[d, "T_x"]\ar[loop right, "T_1", distance=2em]& \\
  & x^3 & x^2 + 1 & x^2 + x &
\end{tikzcd}
\caption{ }\label{fig:model_selection}
\end{subfigure}\caption{\ref{fig:monoid_ab} Monoid over $\{T_x,T_1\}$ with $T_1T_1=T_1$. \ref{fig:model_selection} Action of $T_x, T_1$ on polynomials.}\label{fig:6}
\end{figure}

The example of sparse polynomial regression model selection can be analyzed in terms of program transformations to illustrate our approach.
Let $M$ be a monoid with two generators, $T_x,T_1$ and one equation $T_1\dot T_1 = T_1$. The elements in this monoid are strings of these two symbols and the multiplication operation is concatenation. Define the action of $M$ on the set of \emph{formal polynomials} in $x$ by the action of the generators.

$T_x$ modifies a polynomial by multiplying by $x$, so that $T_x(p) = xp$ and $T_1$ modifies a polynomial by adding a constant term if it doesn't already exist $T_1(p) = p + 1$, where $1+1=1$.
The set of all polynomial regression models lie in the $M$-orbit of the constant polynomial $p(x) = 1$.
Figure~\ref{fig:6} shows how polynomial regression models can be derived as transformations with $M$.
This method of deriving more complex models from simple models can be generalized and implemented in highly generic software.
In our representation of models as categories and model augmentations as functors between those categories, there is always a model transformation monoid.
If that monoid is finitely generated, we can represent the space of possible transformations with a directed graph, analogous to a Cayley graph. And the set of possible models  is the action of the transformation monoid on a \emph{base model}.


This monoidic representation can be further applied to combining models and statistical regression models into workflows using the ideas from Fong and Spivak~\cite{fong2018}.
By representing the workflow as a data structure and the code that defines a workflow as a model, we are able to recursively apply model augmentation and achieve workflow modification using the same tools, now operating on compositions of models.

\paragraph{Conclusions}\label{{conclusions}}As scientific modeling code increases in complexity, it becomes harder for readers to understand all the semantic details that are being represented.
We present a methodology for translating properties of functors to scientifically meaningful relationships between models, and a framework for extracting information from, reasoning about, and augmenting computational models.
We have packaged these ideas into \semanticModels{}, a software package that allows for extension, modification, and composition of models written in standard Julia code.
This library allows model transformations to be defined semantically, and is based on category theoretic knowledge representations, which aims to be practically usable and theoretically sound.

\paragraph{Acknowledgments}
This material is based upon work supported by the Defense Advanced Research Projects Agency (DARPA) under Agreement No. HR00111990008.

\nocite{*}
\bibliographystyle{eptcs}
\bibliography{refs}
\end{document}